\documentclass{PoS}
\usepackage{lineno}
\usepackage{textcomp}
\usepackage{siunitx}

\newcommand{\pt}{\ensuremath{p_{\rm T}}}
\newcommand{\wpm}{\ensuremath{\rm W^{\pm}}}

\title{Measurement of W-boson production in p-Pb collisions at the LHC with ALICE}

\ShortTitle{Measurement of W-boson production in p-Pb collisions at the LHC with ALICE}

\author{\speaker{Kgotlaesele Senosi}\\%
for the ALICE Collaboration\\
 
University of Cape Town\\
Rondebosch, Cape Town, 7700, South Africa \\
iThemba Laboratory of Accelerator Based Sciences\\
Old Faure Road, Faure, South Africa\\
        E-mail: \email{ksenosi@cern.ch}}

\abstract{
ALICE (A Large Ion Collider Experiment) is designed and optimized to study ultra-relativistic heavy-ion collisions, in which a hot and dense strongly-interacting medium is created. W bosons are produced in hard scattering processes occurring at the early stage of the collision and, since they are not affected by the strong interaction, they can be used as a benchmark for medium-induced effects. In proton-nucleus collisions the production of W bosons can be used to study the modification of
parton distribution functions in the nucleus and to test the validity of binary collision scaling. The latter is studied by measuring the yield of W bosons in different intervals of event activity. In ALICE, the production of W bosons is measured via the contribution of their muonic decays to the inclusive $p_{\rm T}$-differential muon yield reconstructed with the muon spectro-meter at forward ($2.03 < \mathit{y}^{\mu}_{cms} < 3.53$) and backward rapidity ($-4.46< \mathit{y}^{\mu}_{cms} <-2.96$). The recent results from p--Pb collisions at a centre-of-mass energy of $\sqrt {s_{\rm NN}}$ = 5.02 TeV are presented and the measured cross sections are compared to perturbative Quantum Chromodynamics calculations at next-to-leading order.
}

\FullConference{53rd International Winter Meeting on Nuclear Physics\\
		 26-30 January 2015\\
		 Bormio, Italy}

\begin{document}

\section{Introduction}
The high energies available in hadronic collisions at the Large Hadron Collider allow for hard probes, for example heavy quarks, quarkonia, high-\pt\ jets and intermediate vector bosons (\wpm\ and $\mathrm{Z^{0}}$), to be produced abundantly. The high mass of W bosons (\wpm) implies that they are formed during initial hard parton scattering processes with a formation time of about 0.003 fm/$c$. The lifetime, which is inversely proportional to their width, is about 0.09 fm/$c$. Precise theoretical predictions of the W-boson cross sections in proton-proton collisions make them good standard candles for luminosity measurements. In addition, they can be used to constrain parton distribution functions (PDFs) at high momentum transfer (\textit{Q}) equal to the mass of the vector bosons. In high-energy heavy-ion collisions where nuclear matter undergoes a phase transition to a deconfined state called Quark-Gluon Plasma (QGP), W bosons will decay either before or during the formation of the QGP. Therefore, W bosons give access to the initial-state properties in nuclear collisions and their yields provide a benchmark for the binary collision scaling of hard processes.
Vector bosons and their leptonic decay products do not interact strongly and thus they should not be affected by the QGP. The isospin dependence of the formation of W bosons and the weak coupling nature makes them good probes to study initial-state effects (for example, the nuclear modification of PDFs and isospin effects) in proton-lead and lead-lead collisions \cite{hannu,ru}. 
\section{Experimental apparatus and data sample}
The ALICE experiment \cite{alice} is an ensemble of various detectors each with a specific purpose. In this analysis the V0 scintillator arrays, zero degree calorimeters (ZNC and ZNA) and the silicon pixel detectors (SPD) are used to provide global information about the event. The V0 arrays, placed on either side of the interaction point, covering the pseudorapidity ranges $2.8 < \eta_{\rm lab}\ < 5.1$ (V0A) and $-3.7 < \eta_{\rm lab}\ < -1.7$ (V0C) are used for triggering as well as for the characterization of events according to their activity. Other detectors used to characterize events according to their activity are the first layer of the SPD (CL1) placed in the central barrel covering the pseudorapidity interval $|\eta_{\rm lab}| < 0.9$ and the zero degree calorimeters (ZNC and ZNA) located 112 meters away on either side of the interaction point along the beam pipe. The SPD is also used to determine the interaction vertex.
The muon spectrometer \cite{alice} which has a forward pseudorapidity coverage of $-4.0 < \eta_{\rm lab}\ < -2.5$ consists of an absorber, five tracking stations, a dipole magnet, a muon filter and two triggering stations. The absorber reduces the background from pions and kaons, such that mostly muons enter the tracking system where their transverse momentum (\pt) is determined. Furthermore, the muon filter, placed between the tracking and trigger stations stops remaining background hadrons and soft muons to ensure high purity of the triggered muon-candidate sample. Furthermore, the triggering stations provide an approximate \pt\ measurement which can be employed for the trigger decision.

The current analysis is based on a data sample collected in proton-lead collisions at $\sqrt{s_{\rm NN}} = 5.02$ TeV ($E_{\rm proton} = 4$ TeV and $E_{\rm lead} = 1.58$ $A$TeV) in which two beam configurations were used, protons going towards the spectrometer (p-Pb) and vice-versa (Pb-p). The asymmetry in energy translates into the centre-of-mass system being boosted by $\Delta y = 0.465$ in the proton direction. The rapidity intervals covered by the muon spectrometer are $2.03 < y_{cms}^{\mu} < 3.53$ (p-Pb) and $-4.46 < y_{cms}^{\mu} < -2.96$ (Pb-p), respectively, where the direction of the proton defines positive rapidities. The integrated luminosities were measured to be 4.9 $\rm nb^{-1}$ and 5.8 $\rm nb^{-1}$ for p-Pb and Pb-p collisions respectively. The data sample consists of events collected with minimum bias (MB) and single-muon triggers. The MB trigger is defined as the coincidence of signals in both V0 detectors and the beam counters, whereas the muon trigger is the coincidence between a MB trigger and a track in the triggering station with $p > $ 4 GeV/\textit{c}. At track level a geometrical acceptance cut ($-4.0 < \eta_{\rm lab}\ < -2.5$) is applied, and the muon track is required to come from an interaction vertex well reconstructed with the SPD. Another geometrical cut taken into consideration is the angle of the tracks at the end of the absorber covering the range ($170\si{\degree} < \theta_{\rm abs} < 178\si{\degree}$) which rejects particles crossing non-uniform material sections at the absorber edges. In order to remove the background from interactions between beam particles and the residual gas, an additional cut is employed based on the product of the muon track momentum and its transverse distance to the interaction point. This cut removes tracks which do not point to the interaction vertex reconstructed with the SPD.
\section{Analysis strategy}
\begin{figure}
\centering
 \includegraphics[width=0.4\textwidth]{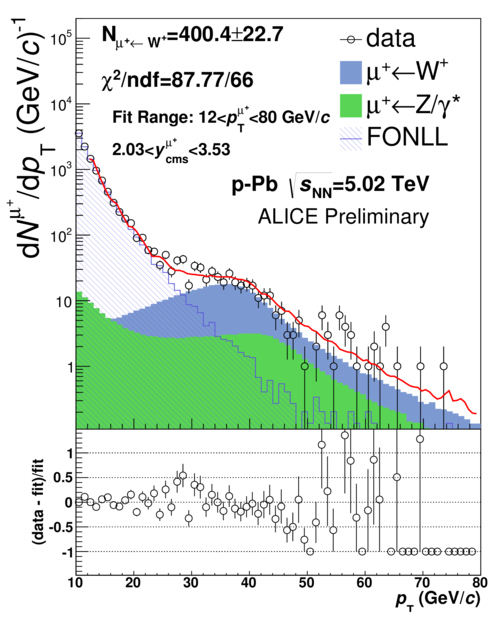}
 \includegraphics[width=0.4\textwidth]{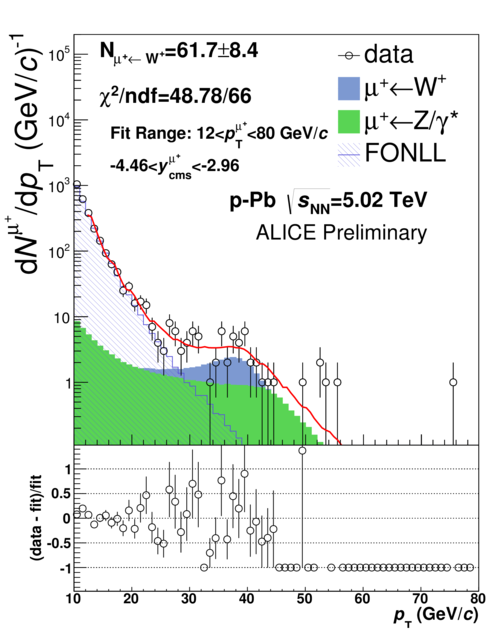}
 \caption{Examples of the combined fit for forward (left) and backward (right) rapidity in the case of positive muons.}
 \label{fits1}
\end{figure}

The analysis is based on the extraction of the W-boson decay contribution to the total single muon transverse-momentum spectrum. The semi-muonic decays of W bosons form a Jacobean peak with a maximum around $\pt\ \sim M_{\rm W}/2$. Muons from Z-boson decays ($\rm Z/\gamma^{*}$) decays represent the dominant background source above $\pt\ \sim 35$ GeV/\textit{c} whereas single muons from the decays of heavy-flavour hadrons dominate the lower \pt\ region ($10 < \pt\ < 35$ GeV/\textit{c}) \cite{zaida}. In order to extract the number of W bosons (N$_{W\leftarrow \mu}$) from the single-muon \pt\ spectrum, a combined fit composed of suitable templates for the various contributions is used. The templates for muons from  W- and Z-boson are based on the next-to-leading-order event generator POWHEG \cite{powheg} and also PYTHIA6.4 \cite{pythia}. The templates are varied in order to estimate the systematic uncertainty from the yield extraction, as well as from detector effects. The templates are generated for pp and pn collisions to take into account the isospin dependence of W- and Z-boson production. These templates are produced in the individual $\eta_{\rm lab}$ intervals. Then they are combined according to:    
\begin{equation}
f_{\rm \mu \leftarrow W,Z^{0}/\gamma^{*}}^{pPb} = a\cdotp\frac{Z}{A}\cdotp f^{pp}_{\rm \mu \leftarrow W,Z^{0}/\gamma^{*}} + a\cdotp\frac{A-Z}{A}\cdotp f^{pn}_{\rm \mu \leftarrow W,Z^{0}/\gamma^{*}}
\end{equation}
where $A=208$, $Z=82$ and $a$ is the normalization factor. The description of background from heavy-flavour (HF) decays is based on perturbative Quantum ChromoDynamics (pQCD) calculations at fixed-order-next-to-leading-log (FONLL) \cite{fonll} and, alternatively, on a phenomenological function inspired by a function previously used by the ATLAS experiment \cite{atlas} at the LHC:
\begin{equation}
 f_{\mu \leftarrow \rm HF}(\pt) = c\cdotp\frac{\exp(-d\cdotp\sqrt{\pt})}{\pt^{2.5}}.
\end{equation}
The three templates are then combined into a fit function
\begin{equation}
\label{comb_fit}
 f(\pt) = N_{\mu \leftarrow \rm HF} \cdot f_{\mu \leftarrow \rm HF} + N_{\mu \leftarrow \rm W}\cdot f_{\mu \leftarrow \rm W} + N_{\mu \leftarrow {\rm Z^{0}}/\gamma^{*}}\cdot f_{\mu \leftarrow {\rm Z^{0}}/\gamma^{*}}
\end{equation}
where $f_{\mu \leftarrow \rm HF}(\pt)$ is either the phenomenological function or the FONLL-based heavy-flavour decay-muon template, and $f_{\mu \leftarrow \rm W}$ and $f_{\mu \leftarrow {\rm Z^{0}}/\gamma^{*}}$ are W- and $\rm Z^{0}/\gamma^{*}$-boson decay templates. In Equation \ref{comb_fit}, $N_{\mu \leftarrow \rm HF}$ and $N_{\mu \leftarrow \rm W}$ are free parameters whereas $N_{\mu \leftarrow {\rm Z^{0}}/\gamma^{*}}$ is constrained by the ratio $N_{\mu \leftarrow {\rm Z^{0}}/\gamma^{*}}/N_{\mu \leftarrow \rm W}$ which is computed from cross sections obtained with POWHEG. An example of this combined fit is shown in Figure \ref{fits1}. The number of W bosons extracted from this fit is affected by the uncertainties in the description of the heavy-flavour decay muon background as well as the knowledge of the ratio of Z to W bosons. The systematic uncertainty on this
ratio is evaluated using PYTHIA6.4 \cite{pythia} calculations. A source of systematic uncertainty related to detector effects is determined by using the templates with different alignment configurations. Uncertainties on the signal extraction, which include alignment effects and also the stability with the variation of the fit range, vary between 6\% and 24\%. The modification of nuclear PDFs, which is not included in the POWHEG templates, is taken into account by the EPS09 \cite{eps09} parametrization included in the PYTHIA6.4 templates. 
The yield of muons from W-boson decays is computed as a weighted average over all trial fits where the statistical uncertainties are propagated per fit, and the systematic uncertainties are computed as the variance of the yield of muons obtained from these fits. 

\section{Results}
\begin{figure}
\centering
 \includegraphics[width=0.7\textwidth]{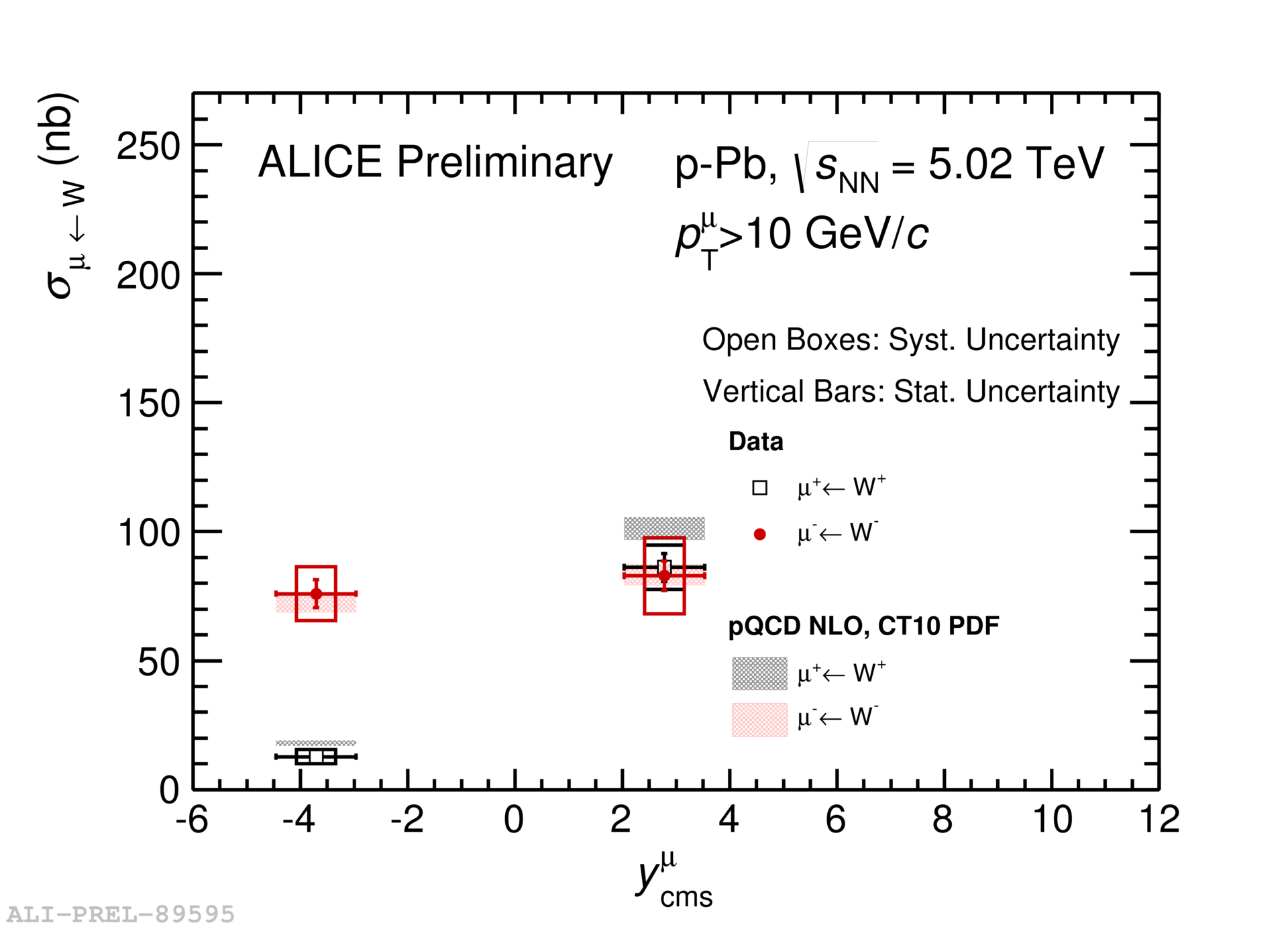}
 \includegraphics[width=0.7\textwidth]{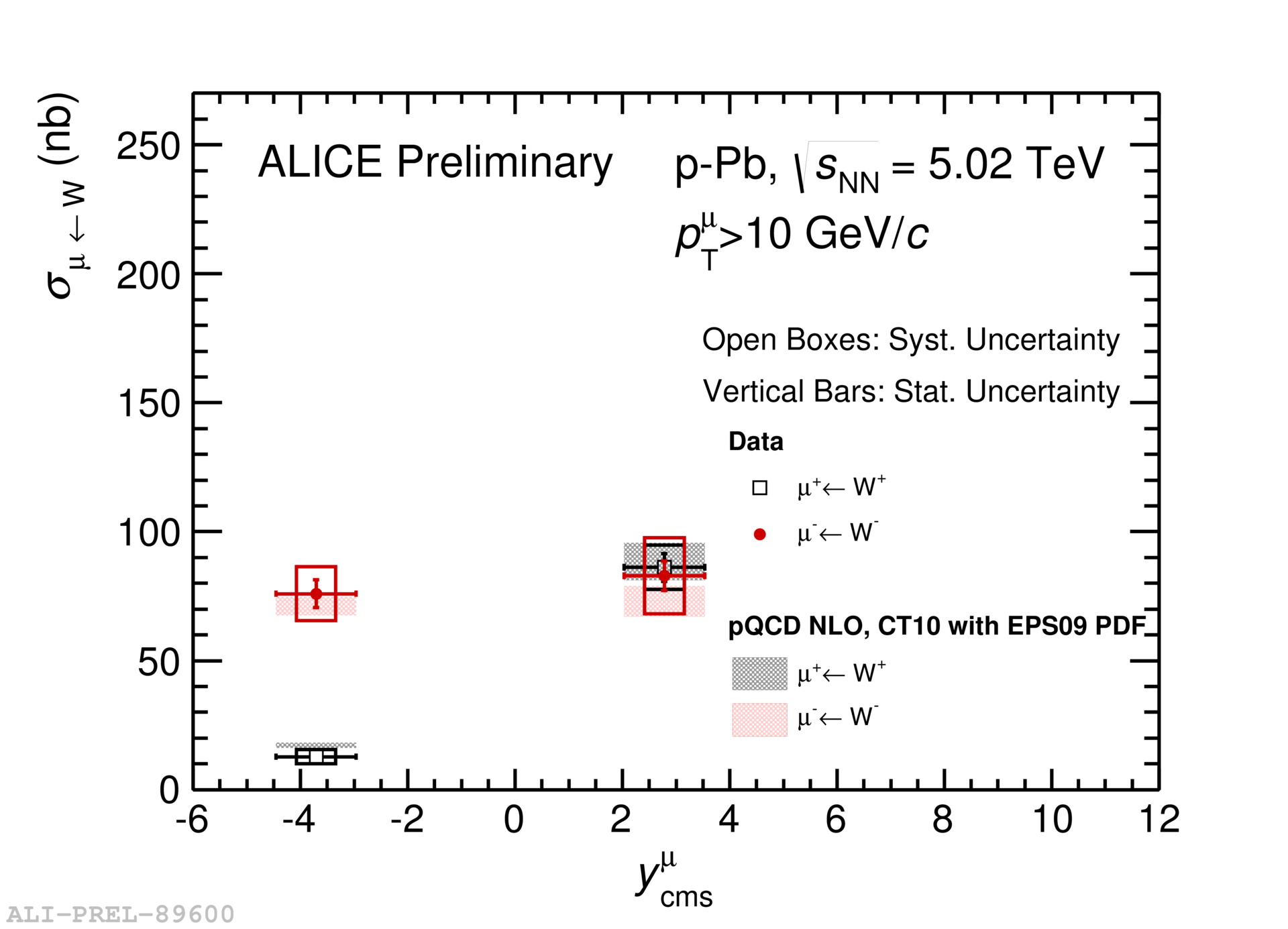}
 \caption{The measured cross section of muons from W-boson decays at forward and backward rapidity compared with pQCD calculations with CT10 PDFs (top) and CT10 PDFs including EPS09 nuclear shadowing (bottom).}
 \label{cross_section}
\end{figure}

\begin{figure}
\centering
 \includegraphics[width=0.48\textwidth]{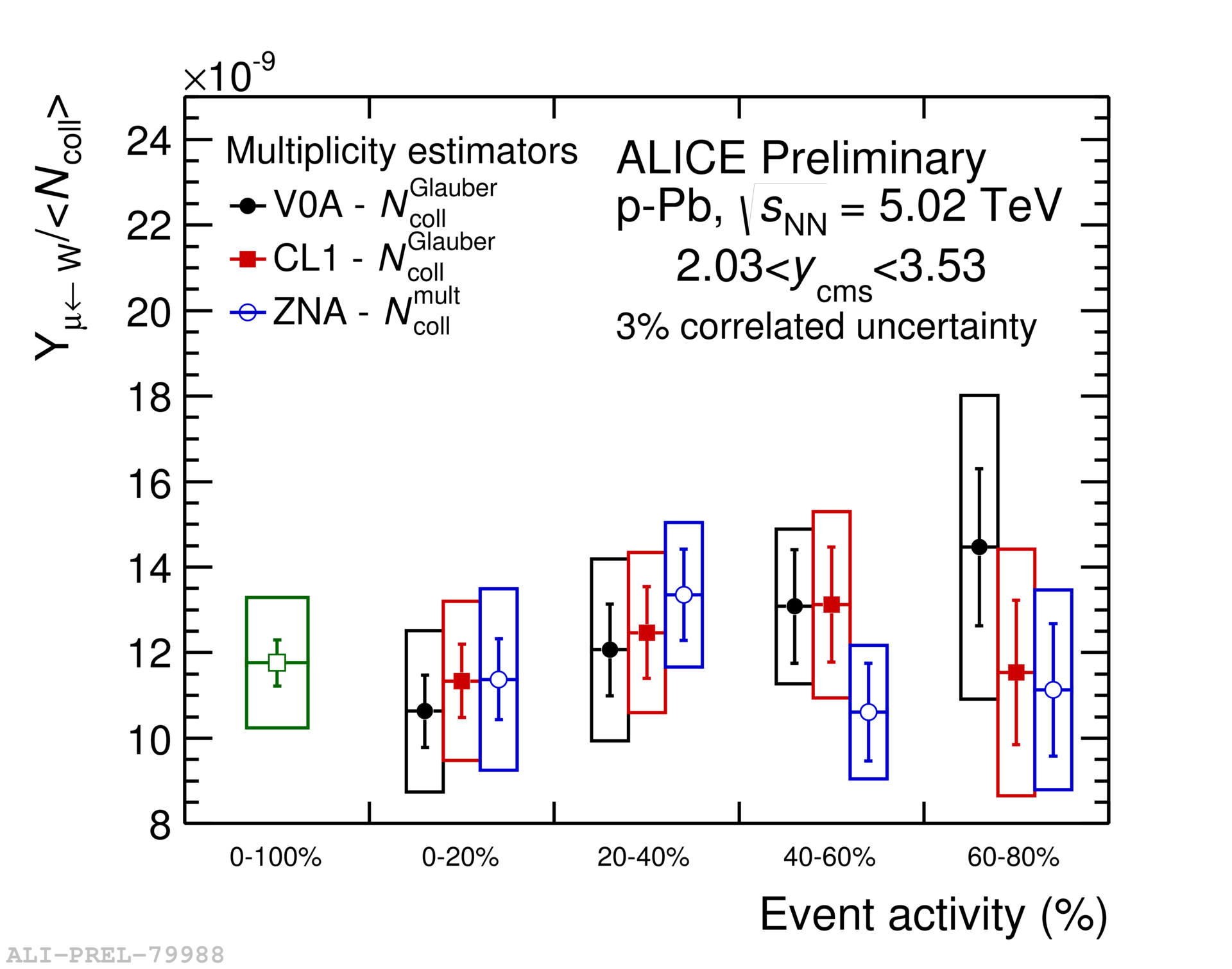}
 \includegraphics[width=0.48\textwidth]{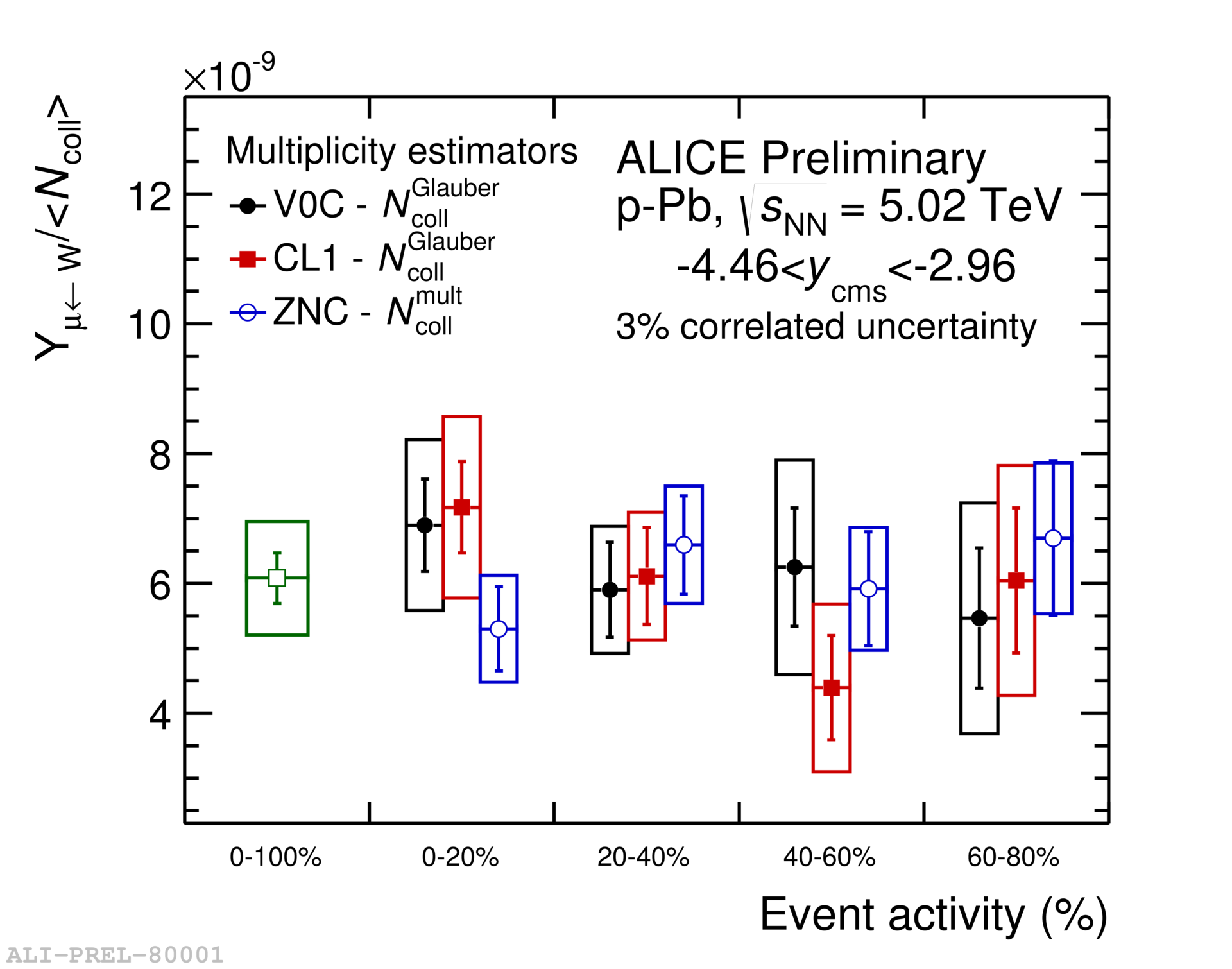}
 \caption{W-boson yield normalized to $\langle N_{\rm coll}\rangle$ shown as a function of event-activity.}
 \label{binary_scaling}
\end{figure}

The $\rm W^{+}$- and $\rm W^{-}$-boson cross sections are computed by correcting their yields for acceptance and efficiency, and then normalizing to the related integrated luminosity. The measured cross sections of W bosons decaying to muons ($\mu^{\pm} \leftarrow \mathrm{W}^{\pm}$) in the spectrometer acceptance are compared with pQCD predictions with CT10 PDFs \cite{ct10} and with CT10 PDFs modified with the EPS09 parametrization \cite{eps09} of nuclear shadowing as shown in Figure \ref{cross_section}. The binary scaling of W-boson yields is tested by dividing the measured yields of muons from W-boson decays (sum of separate N$_{\mu^{+}\leftarrow \mathrm{W}^{+}}$ and N$_{\mu^{-}\leftarrow \mathrm{W}^{-}}$) per event by the average number of binary nucleon-nucleon collisions $\langle N_{\rm coll}\rangle$ \cite{centrality} which characterizes the event activity. This measurement is done for minimum-bias events (no event activity selection) and in classes of event activity. $\langle N_{\rm coll}\rangle$ is obtained from Glauber-Monte-Carlo fits to the V0A amplitude and the number of clusters in CL1, whereas for the ZN estimator the assumption made is that the number of particles at midrapidity is correlated with the number of nucleons participating in the collision \cite{centrality}. Figure \ref{binary_scaling} shows the yield normalized to $\langle N_{\rm coll}\rangle$ for both backward and forward rapidity. The uncertainties on the ratio Yield$_{\mu \rightarrow \mathrm{W}}$/$<N_{\mathrm{coll}}>$ includes the uncertainties on $\langle N_{\mathrm{coll}}\rangle$ which varies between 8\% and 21\% depending on the event-activity class.   

\section{Conclusions}
The production cross sections of W bosons in p-Pb collisions at $\sqrt{s_{\mathrm{NN}}} =$ 5.02 TeV have been measured in two rapidity intervals. The comparison of these cross sections with theoretical predictions based on a pQCD calculation with CT10 PDFs shows agreement within uncertainties. Taking into account the EPS09 prescription of nuclear shadowing of the PDFs further improves the agreement between the calculation and the data at forward rapidity where shadowing is expected to be important. 
The measured yield of muons from W-boson decays normalized to $\langle N_{\mathrm{coll}}\rangle$ as function of event-activity shows that the W-boson production scales with $\langle N_{\mathrm{coll}}\rangle$.


\begin{thebibliography}{99}
\bibitem{hannu}
  Paukkunen H and Salgado C \emph{JHEP} \textbf{1103} (2011) 071
  
  \bibitem{ru}
  Ru P, Zhang B, Cheng L, Wang E and Zhang W \emph{arXiv}:1412.2930 (2014)
  
  \bibitem{alice}
  Aamodt K \emph{et al}. (ALICE Collaboration) 2008 JINST \textbf{3} S08002
 
  \bibitem{zaida}
  Conesa del Valle Z \emph{Eur. Phys. J. C}\textbf{61} (2009) 729
  
  \bibitem{powheg}
  Alioli S, Nason, P, Oleari C and Re E \emph{JHEP} \textbf{0807} (2008) 060 
  
  \bibitem{fonll}
  Cacciari M, Greco M and Nason P \emph{JHEP} \textbf{1210} (2012) 137
  
  \bibitem{atlas}
  ATLAS Collaboration, ATLAS-CONF-2011-078
  
  \bibitem{pythia}
  Sjostrand T, Mrenna S and Skands P \emph{JHEP} \textbf{05} (2006) 026
  
  \bibitem{eps09}
  Eskola KJ, Paukkunen H and Salgado CA \emph{JHEP} \textbf{0904} (2009) 065
  
  \bibitem{ct10}
  Lai HL \emph{et al}. \emph{Phys. Rev. D} \textbf{82} (2010) 074024
  
  \bibitem{centrality}
  Toia A \emph{et al}.(ALICE Collaboration) \emph{Nucl. Phys. A} \textbf{926} (2014) 78-84
\end{thebibliography}
\end{document}